\documentclass[amsmath,amssymb,prx,aps,showpacs,twocolumn,10pt]{revtex4-2}
\usepackage{amsmath,amsfonts,amssymb,amsthm,graphics,graphicx,epsfig,bbm}
\usepackage[colorlinks=true,citecolor=blue,linkcolor=red,urlcolor=blue]{hyperref}
\usepackage[usenames]{color}
\usepackage{graphicx}
\usepackage{subfigure}
\usepackage{amsmath}
\usepackage{epsfig}
\usepackage{dcolumn}
\usepackage{bm}
\usepackage{color}
\usepackage{epstopdf}
\usepackage{amssymb}
\usepackage{amstext}
\usepackage{latexsym}
\usepackage{hyperref}
\usepackage{amsfonts}
\usepackage{psfrag}
\usepackage{xcolor}
\usepackage[normalem]{ulem}
\usepackage{dsfont}
\usepackage{txfonts}
\usepackage{overpic}
\usepackage{ifthen}
\usepackage{amsthm}

\usepackage{youngtab}

\usepackage{amsmath, amssymb, amsthm}

\newcommand{\an}[2]{\ifthenelse{\equal{#1}{}}{\ensuremath{\hat{#1}_{#2}}}{\ensuremath{\hat{#1}^{\protect\phantom{\dagger}}_{#2}}}}

\begin{document}

\title{Complexification of Quantum Signal Processing and its Ramifications}

\author{V. M. Bastidas$^{1,2}$}
\email{victor.bastidas@ntt-research.com}

\author{K. J. Joven$^1$}
\email{kevin.joven@ntt-research.com}
\affiliation{%
$^1$Physics and Informatics Laboratory, NTT Research, Inc., 940 Stewart Dr., Sunnyvale, California, 94085, USA
} 
\affiliation{%
$^2$Department of Chemistry, Massachusetts Institute of Technology, Cambridge, Massachusetts 02139, USA
} 

\date{\today}

\begin{abstract}

In recent years there has been an increasing interest on the theoretical and experimental investigation of space-time dual quantum circuits. They exhibit unique properties and have applications to diverse fields. Periodic space-time dual quantum circuits are of special interest, due to their iterative structure defined by the Floquet operator. A very similar iterative structure naturally appears in Quantum Signal processing (QSP), which has emerged as a framework that embodies all the known quantum algorithms. However, it is yet unclear whether there is deeper relation between these two apparently different concepts.
In this work, we establish a relation between a circuit defining a Floquet operator in a single period and its space-time dual defining QSP sequences for the Lie algebra sl$(2,\mathbb{C})$, which is the complexification of su$(2)$. First, we show that our complexified QSP sequences can be represented in terms the Lorentz group acting on density matrices and that these sequences can be interpreted as hybrid circuits involving unitaries and measurements. We also show that unitary representations of our QSP sequences exist, although they are infinite-dimensional and defined for bosonic operators in the Heisenberg picture. Finally, we also show the relation between our complexified QSP and the nonlinear Fourier transform for sl$(2,\mathbb{C})$, which is a generalization of the previous results on su$(2)$ QSP.

\end{abstract}

\maketitle
\section{Introduction \label{SecI}}
In recent years, the field of space-time dual quantum circuits has emerged as a link between quantum information science, quantum signatures of chaos, statistical and condensed matter physics~\cite{Piroli2020,fisher2023random, Claeys2021,Ippoliti2022}. The space-time duality appears in the dynamics of the kicked one-dimensional quantum Ising model~\cite{Akila2016,Braun2020,Bertini2018} and quantum circuits~\cite{BertiniNew2019,Fritzsch2021}. The field of space-time dual quantum circuits has been extensively developed in a number of different directions such as dual circuits in random geometries~\cite{kasim2023dual},  measurement induced phase transitions~\cite{Lu2021}, measurement-based quantum computation~\cite{Stephen2024}, crystalline quantum circuits~\cite{Sommers2023} and to investigate the sensitivity to the initial and boundary conditions using out-of-time-order correlators~\cite{PhysRevResearch.4.L022039}. 
As might be expected, space-time quantum circuits can also be interpreted as quantum algorithms and it is interesting to use tools of quantum information science to investigate them~\cite{Ippoliti2021,Aravinda2021,Foligno2023,Bertini2022}. This perspective can benefit both communities and also provide valuable insights for other fields.

Quantum Signal Processing (QSP) is one of the most important developments in the field of quantum algorithms in recent years~\cite{Low2017,Martyn2021,Sinanan2023}. Its fascinating mathematical structure allows one to effectively unify all the known quantum algorithms within the same framework~\cite{Martyn2021}. Further, QSP provides a unique perspective on how to find new quantum algorithms by mapping problems in quantum computation to the field of functional analysis and algebraic geometry~\cite{Rossi2022, berntson2024}. On top of that, QSP methods such as qubitization and block encoding are currently widely used in different fields ranging from quantum simulation of chemistry problems to the development of quantum algorithms~\cite{Zeytinouglu2022,Laneve2023}.

In its original form, the QSP theorem provides us with a relation between a finite sequence qubit operations generated by the su$(2)$ algebra and polynomial approximations of a response function~\cite{Low2017,Martyn2021}. A recent work has extended the QSP theorem to sequences of operations generated by the lie algebras  su$(1,1)$~\cite{rossi2023}, su$(N)$~\cite{Laneve2023}, and for the Onsager algebra appearing in the solution of the quantum Ising model~\cite{Bastidas2024}. In the case of su$(1,1)$, it was shown that there is no finite dimensional unitary representation of the QSP sequence~\cite{rossi2023}. For this reason, the only way to define it is to work with the squeezing algebra and continuous variables~\cite{rossi2023}. Recently, it has been shown that QSP sequences for su$(2)$ and su$(1,1)$ are intimately related to the nonlinear Fourier transformation~\cite{Gardner1967,Zakharov1972,ablowitz1974,tao2012nonlinear}, which is the nonlinear version of the well-known Fourier transform. In despite all the recent advances in the field, it is still an open question whether there is a more general underlying mathematical structure that relates the QSP theorems for su$(2)$ and su$(1,1)$. 

\begin{figure}
    \includegraphics[width=0.45 \textwidth]{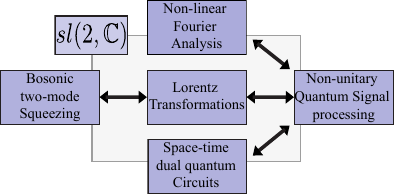}
    \caption{Scheme of the multiple relations between quantum signal processing with  sl$(2,\mathbb{C})$ and other fields discussed in this work. 
    }
    \label{Fig3}
\end{figure}

In our work, we demonstrate the intimate relation between space-time dual periodic quantum circuits and quantum signal processing with the sl$(2,\mathbb{C})$ Lie algebra~\cite{mcmahon1975sl,Polyzou2019}, which is the complexification of  su$(2)$ and su$(1,1)$. We show that a quantum circuit with $L_x=N$ qubits and a single time step $L_t=1$ maps to a QSP sequence with complex parameters that can be directly written using generators of sl$(2,\mathbb{C})$. As this algebra plays a very important role in special relativity~\cite{mcmahon1975sl,Polyzou2019}, we define a non-unitary QSP sequence acting on density matrices that is in direct correspondence to the action of the Lorentz group~\cite{Teodorescu2003}. Similarly to su$(1,1)$, we demonstrate that QSP sequences with the sl$(2,\mathbb{C})$ algebra  have infinite dimensional unitary representations intimately related to squeezing. Motivated by a representation found by Dirac, we show that the unitary representation is related to a bosonic representation of the Lorentz group using phase shifters, beam splitters, and single-mode squeezing operations. Finally, we show that QSP sequences with  sl$(2,\mathbb{C})$ are directly related to the complex discrete Fourier transform and we provide an intuitive and geometric picture of this relation.

The paper is organized as follows. In section~\ref{SecII}, we discuss the motivation of our work and the relation between space-time dual quantum circuits and QSP sequences with  sl$(2,\mathbb{C})$ algebra. After that, in section~\ref{SecIII}, we consider in more detail the algebraic properties of  QSP sequences with  sl$(2,\mathbb{C})$  and the relation to the Lorentz group. The non-unitary representation of QSP sequences for sl$(2,\mathbb{C})$ is constructed in section~\ref{SecIV}, and the corresponding infinite-dimensional representation is discussed in section~\ref{SecV}. Finally, we explore the relation to the non-linear Fourier transform in section~\ref{SecVI}. As a guide for the reader, Fig.~\ref{Fig3} shows a diagram that summarizes the concepts and fields that we analyze in our work.

\section{Motivation: Dual Quantum circuits and Quantum signal processing with the Ising model\label{SecII} } 

Let us start by considering the dynamics of a periodic quantum circuit described by the following Floquet operator~\cite{Lu2021}
\begin{align}
          \label{eq:QuantumCircuit}
\hat{U}_{\text{F}}=e^{\mathrm{i}\sum^{N}_{j=1} \alpha_{j} X_j}e^{\mathrm{i}  \sum^{N}_{j=1} ( \theta Z_j Z_{j+1}+\phi_j Z_j )}
\ ,
\end{align}
where $X_j$, $Y_j$, and $Z_j$ denote the Pauli operators at the $l$-th site. The chain is in a random transverse $\alpha_{j}$ and longitudinal $\phi_{j}$ fields, while $\theta$ is proportional to the Ising interaction strength. The Floquet operator is in fact the evolution operator within a period $T$ of the drive for fixed random instance of the disorder~\cite{Sunderhauf2018}. Thus, if one thinks in terms of space and time, the Floquet operator propagates a spatial lattice with $L_x=N$ sites along $L_t=1$ time step in time. 

\begin{figure}
\centering
    \includegraphics[width=0.49 \textwidth]{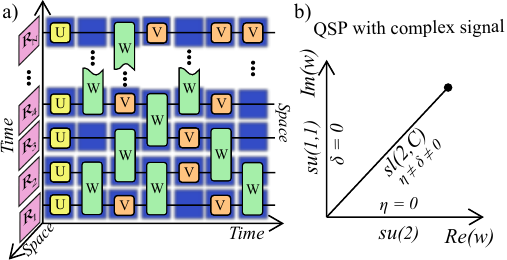}
    \caption{Space-time duality and quantum signal processing. a) Illustrates a quantum circuit with $L_x=N$ qubits and a single time step $L_t=1$ and its space-time dual that corresponds to a circuit with a single qubit $L_x=1$ and $L_t=N$ time steps. The fundamental operations that define the circuit are denoted by $\hat{U}_j=e^{\mathrm{i} \alpha_{j} X_j}$, $\hat{W}_j=e^{\mathrm{i}  \theta Z_j Z_{j+1}}
$ and $\hat{V}_j=e^{\mathrm{i}  \phi_j Z_j }
$. b) The dual circuit corresponds to a quantum signal processing (QSP) sequence with complex signal $w=\tilde{\theta}/2=(\delta+\mathrm{i} \eta)/2$ and operators  $\hat{R}_r=e^{\mathrm{i} \tilde{\theta} X_r }e^{\mathrm{i} \phi_{r} Z_r}$. The operations of the sequence belong to the group sl$(2,\mathbb{C})$. When the signal is real $( \eta=0)$, we recover the usual su$(2)$ QSP, while if the signal is just imaginary $( \delta=0)$ we obtain QSP for su$(1,1)$.
    }
    \label{Fig1}
\end{figure}

Recently, there is an onset of interest on space-time dual quantum circuits. In the dual picture, the quantum circuit in Eq.~\eqref{eq:QuantumCircuit} can be replaced by a quantum circuit that propagates a lattice with a single site $\tilde{L}_x=1$ over $\tilde{L}_t=N$ time steps, as follows
\begin{align}
          \label{eq:DualQuantumCircuit}
\hat{U}_{\text{DF}}=\prod^{N}_{r=1} 
e^{\mathrm{i} \tilde{\theta} X_r}e^{\mathrm{i} \phi_r Z }
 \ ,
\end{align}
where $\tan\tilde{\theta}=-\mathrm{i}e^{-2\mathrm{i}\theta}$~\cite{Lu2021}. The dual unitary strongly resembles a QSP sequence. Importantly, the phases $\phi_{j} $ defining the unitary operators $\hat{V}_j=e^{\mathrm{i} \phi_{j} Z_j}$ of the original lattice can be interpreted as a QSP sequence in time with angles $\{\phi_r\}$  in the dual picture, as it is illustrated in Fig.~\ref{Fig1}~a).  

The previous discussion is the core of our work. We shall investigate QSP sequences of the canonical form in Eq.~\eqref{eq:DualQuantumCircuit} where the signal is a complex parameter. In the next section, we will explore the algebraic interpretation of this problem in terms of sl$(2,\mathbb{C})$, which is the complexification of the Lie algebra su$(2)$.

\section{Quantum signal processing with sl$(2,\mathbb{C})$ \label{SecIII} } 
In the previous section we encounter a situation where a dual quantum circuit looks like a QSP sequence with a complex signal $\widetilde{\theta}$. As a result, the QSP sequence is non-unitary. Motivated by this, in this section we investigate non-unitary QSP sequences with the canonical form
\begin{align}
\label{eq:ComplexQSPDef}
\hat{V}_{\vec{\phi}}(w)&=e^{\mathrm{i} \phi_0 Z}\prod^{2d+1}_{r=1}e^{\mathrm{i} w X }e^{\mathrm{i} \phi_r Z}  
\nonumber\\
&=\begin{bmatrix}
    P(\cos w) &  \mathrm{i} Q(\cos w) \sin w  \\
     \mathrm{i}R(\cos w) \sin w &S(\cos w)  \\
    \end{bmatrix}
\ ,
\end{align}
where $\vec{\phi}=(\phi_0,\phi_1,\dots,\phi_{2d})$ and $
P(\cos w)S(\cos w) +R(\cos w)Q(\cos w)\sin^2 w=1$. In fact, the QSP sequence turns out to be an element of the Lie group SL$(2,\mathbb{C})$ of $2\times 2$ complex unimodular matrices ($2\times 2$ complex matrices $A$ such that $\det A=1$). In contrast to the original QSP theorem, here the signal is a complex number $w=(\delta+\mathrm{i} \eta)/2$ with $\delta$ and $\eta$ being real. One important aspect here is that ${X,Y,Z}$ are Pauli matrices generating the su$(2)$ Lie algebra as long as one consider a linear combination of Pauli matrices with real coefficients. However, once we allow for complex numbers, we need to work with the complexification of su$(2)$ algebra which is known as sl$(2,\mathbb{C})$. 

The QSP sequence discussed above is non-unitary and can be considered as an analytical continuation of the original QSP~\cite{Low2017,Martyn2021} from the real line to the complex plane. One interesting aspect of sl$(2,\mathbb{C})$ is that it is the complexification of both su$(2)$ as well su$(1,1)$ 
~\cite{lempert1997problem}, which are related to  $\eta$ and $\delta$ defining the signal complex $w$. We can get one or the other by setting $\eta=0$ and $\delta=0$. The sl$(2,\mathbb{C})$ QSP arise when considering $\eta\neq0$ and $\delta\neq 0$ as we depict in Fig.~\ref{Fig1}~b).

For example, if we now consider $\delta=0$, we obtain su$(1,1)$ QSP recently developed~\cite{rossi2023}. In this case, as $\cos (\mathrm{i}\eta)=\cosh \eta$ and $\sin (\mathrm{i}\eta)=\mathrm{i}\sinh \eta$ we obtain 
\begin{align}
\label{eq:ImQSPDef}
\hat{V}_{\vec{\phi}}(w)&=e^{\mathrm{i} \phi_0 Z}\prod^d_{r=1} e^{- \eta X }e^{\mathrm{i} \phi_r Z} 
\nonumber\\
&=\begin{bmatrix}
    P(\cosh \eta) & - Q(\cos w) \sinh \eta  \\
     -Q^*(\cosh \eta) \sinh \eta & P^*(\cosh \eta)  \\
    \end{bmatrix}
\ .
\end{align}
Before discussing aspects of this Lie algebra sl$(2,\mathbb{C})$, it is important to discuss the polar decomposition of the Lie group SL$(2,\mathbb{C})$ ~\cite{Polyzou2019}
\begin{align}
          \label{eq:SLGroup}
A=\begin{bmatrix}
    a & b  \\
     c & d  \\
    \end{bmatrix}
    =\exp(\gamma\boldsymbol{m}\cdot\boldsymbol{\sigma})\exp(\mathrm{i}\lambda\boldsymbol{n}\cdot\boldsymbol{\sigma})
\ ,
\end{align}
where $\gamma$ and $\lambda$ are real, $ad-bc=1,$ $\boldsymbol{m}=(m_x,m_y,m_z)$, $\boldsymbol{n}=(n_x,n_y,n_z)$ and $\boldsymbol{\sigma}=(X,Y,Z)=(\sigma_1,\sigma_2,\sigma_3)$. 
 One important aspect to keep in mind is that in the polar decomposition $\exp(\gamma\boldsymbol{m}\cdot\boldsymbol{\sigma})$ is positive definite while $\exp(\mathrm{i}\lambda\boldsymbol{n}\cdot\boldsymbol{\sigma})$ is unitary.

\begin{figure}
    \includegraphics[width=0.49 \textwidth]{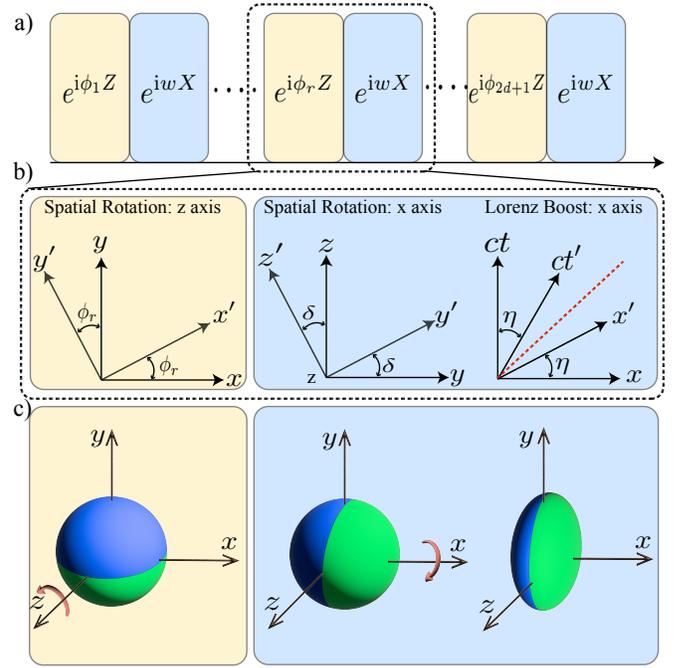}
    \caption{Space-time quantum signal processing. a). Depicts a sl$(2,\mathbb{C})$ quantum signal processing (QSP) sequence for a given complex signal $w=(\delta+\mathrm{i} \eta)/2$. b) Shows the isomorphism between SL$(2,\mathbb{C})$ group and the Lorentz transformations in correspondence to the QSP sequence depicted in a). The qubit rotation $e^{\mathrm{i} \phi_r Z}$ maps to a rotation of the spatial coordinates along the $z$ axis. The real part of the signal $\delta$ produces a spatial rotation along the $x$ axis, while its imaginary part $\eta$ produces a Lorentz boost. That is a space-time rotation preserving the light cone structure represented by the red dotted line. c) Illustrates the action of our QSP sequence on the Bloch sphere. The Lorentz boost effectively squeezes the sphere along the $x$ axis, which is nothing but the relativistic length contraction of an object in a reference frame moving  along the $x$ axis. 
    }
    \label{Fig2}
\end{figure}

Next, let us briefly summarize the main aspects of the Lie algebra sl$(2,\mathbb{C})$. Due to the condition of the determinant being equal to one, the Lie algebra sl$(2,\mathbb{C})$ is six-dimensional and its generated by  $2\times 2$ traceless matrices. This generators are explicitly given by two groups of three generators~\cite{jaffe2013lorentz}. The first one is given by Hermitian matrices

\begin{align}
\label{eq:HermitianGenerators}
J_1=\frac{1}{2}X, \ J_2=\frac{1}{2}Y, \ \text{and} \ J_3=\frac{1}{2}Z
\ .
\end{align}
The second group is composed by anti-Hermitian matrices

\begin{align}
\label{eq:AntiHermitianGenerators}
K_1=\frac{\mathrm{i}}{2}X, \ K_2=\frac{\mathrm{i}}{2}Y , \ \text{and} \ K_3=\frac{\mathrm{i}}{2}Z
\ .
\end{align}
In terms of these generators, we can write our sl$(2,\mathbb{C})$ QSP sequence  in Eq~\eqref{eq:ComplexQSPDef} as
\begin{align}
\label{eq:LorentzQSPDef}
\hat{V}_{\vec{\phi}}&=\prod^{2d+1}_{r=1} e^{2\mathrm{i} \phi_r J_3} e^{\mathrm{i} \delta J_1 + \mathrm{i}\eta K_1}
\end{align}
as we depict in Fig.~\ref{Fig2}~a).

Interestingly, the $6$ aforementioned matrices satisfy the same algebraic relations defining the Lie algebra of the Lorentz group in special relativity~\cite{Bacskal2021,Polyzou2019}
\begin{align}
\label{eq:Lorentz}
[J_i, J_j]&=\mathrm{i}\epsilon_{i,j,k}J_k \ ,
\nonumber\\
 [J_i, K_j]&=\mathrm{i}\epsilon_{i,j,k}K_k \ ,
 \nonumber\\
  [K_i, K_j]&=-\mathrm{i}\epsilon_{i,j,k}J_k
\ .
\end{align}

The operators $J_j$ for $j=1,2,3$ generate spatial rotation along the $x,y$ and $z$ axis, respectively. Similarly, $K_k$ with  $k=1,2,3$ are the Lorentz boost (space-time rotations along $x,y$ and $z$ axis preserving the velocity of light). It is important to have a geometrical image of our QSP sequence in Eq.~\eqref{eq:LorentzQSPDef} in terms of the Lorentz transformations, as we depict in Fig.~\ref{Fig2}~b). Notably, as the signal is complex, the signal operator corresponds to the combined effect of a spatial rotation $e^{\mathrm{i} \delta J_1}
$ and a boost $e^{\mathrm{i}\eta K_1}
$ along the $x$ axis. Correspondingly, the signal processing operator represents a spatial rotation $e^{2\mathrm{i} \phi_r J_3} $ along the $z$ axis.

The QSP sequence for the SU$(2)$ group can be visualized as a quantum walk in the three-dimensional sphere $S^3$~\cite{rossi2023}. However, our non-unitary QSP is determined by six parameters defining a non-compact six-dimensional manifold~\cite{Bacskal2021}, and it is not direct to visualize it as a quantum walk.

We were initially motivated by space-time dual quantum circuits. However, one important question to explore next is whether there is a physical implementation of this algebra that is relevant for other physical systems. This is the purpose of the next section.

\section{Non-unitary Quantum signal processing acting on density matrices and relation to the Lorentz transformations \label{SecIV} } 
From the precious section we learned that QSP with sl$(2,\mathbb{C})$ leads to a non-unitary form of QSP that is intimately related to the algebra of the Lorentz group in special relativity~\cite{Polyzou2019}. Due to its non-unitary character, it makes sense to think of this QSP sequence of as acting on matrices of the form~\cite{Teodorescu2003}
\begin{align}
         \label{eq:QuadrivectorDensityMatrix}
                  \hat{\Omega}_0=\frac{1}{2}\sum_{\nu=0} x_{\nu}\sigma_{\nu}=\frac{1}{2}\begin{bmatrix}
    x_0+x_3 & x_1-\mathrm{i}x_2  \\
    x_1+\mathrm{i}x_2 & x_0-x_3   \\
    \end{bmatrix}
  \ ,
\end{align}
where $x_{\nu}$ are real parameters and $\sigma_{0}=\hat{1}$, $\sigma_{1}=X$, $\sigma_{2}=Y$ and $\sigma_{3}=Z$. The determinant of this density matrix reads $\det\hat{\Omega}_0=(x_0^2-x_1^2-x_2^2-x_3^2)/4$, which is a relativistic invariant in special relativity.

In order to have a better geometrical representation of the action of our QSP sequence, let us assume that the density matrix $\hat{\Omega}_0$ represents a pure state of the form  $\hat{\Omega}_0=(\hat{1}+\boldsymbol{l}\cdot\boldsymbol{\sigma})/2$ with $x_0=1$ and $\boldsymbol{l}=(x_1,x_2,x_3)$ such that $|\boldsymbol{l}|^2=1$. This automatically restricts the spatial coordinates to lie on the sphere $x_1^2+x_2^2+x_3^2=1$ and fixes the value $\det\hat{\Omega}_0=0$ for our relativistic invariant. This defines the Bloch sphere depicted in the left panel of Fig.~\ref{Fig2}~c).

Next, we define the action of our QSP sequence on the matrix $ \hat{\Omega}_0$ as follows
\begin{align}
         \label{eq:QSPActionDensityMatrix1}
                 \hat{\Omega}= \hat{V}_{\vec{\phi}}\hat{\Omega}_0 {\hat{V}_{\vec{\phi}}}^{\dagger}=\frac{1}{2}\sum_{\nu=0} x_{\nu}\hat{V}_{\vec{\phi}}\sigma_{\nu}\hat{V}^{\dagger}_{\vec{\phi}}
                    \ ,
                    \end{align}
where the resulting density matrix $ \hat{\Omega}=\frac{1}{2}\sum_{\nu=0} x'_{\nu}\sigma_{\nu}$ encodes a new set of space-time coordinates $x'_{\nu}$. It is important to remark that this is the usual spinor representation of the Lorentz group~\cite{Polyzou2019}. Next, let us investigate the effect of our QSP sequence on the density matrix in detail. To do this, we multiply  both sides with $\sigma_{\nu}$ and and take the trace to obtain
\begin{align}
         \label{eq:QSPActionDensityMatrix2}
              x'_{\mu}=  \text{Tr}(\sigma_{\mu} \Omega)&=\frac{1}{2}\sum_{\mu=0} x_{\nu} \text{Tr}\left(\sigma_{\mu}\hat{V}_{\vec{\phi}}\sigma_{\nu}\hat{V}^{\dagger}_{\vec{\phi}}\right)=\sum_{\mu=0} \Lambda_{\mu,\nu} x_{\nu}
                    \ ,
\end{align}
where we have defined $2\Lambda_{\mu,\nu}=\text{Tr}\left(\sigma_{\mu}\hat{V}_{\vec{\phi}}\sigma_{\nu}\hat{V}^{\dagger}_{\vec{\phi}}\right)$. Thus, we can clearly see that our QSP sequence acts as Lorentz tranformation $ x'_{\mu}=\sum_{\mu=0} \Lambda_{\mu,\nu} x_{\nu}
$ on the four-vector $x_{\nu}$ encoded in the density matrix as we illustrate in Fig.~\ref{Fig2}~c). This transformation leaves the determinant of the matrix $\det\hat{\Omega}=\det\hat{\Omega}_0=0$ unchanged but it does not preserve its trace because $\text{Tr}\ \Omega= x'_{0}=\sum_{\mu=0} \Lambda_{0,\nu} x_{\nu}
$. Note that while the spatial rotations preserve the trace, the Lorentz boosts effectively squeeze the Bloch sphere along a given direction as we show in the right panel of Fig.~\ref{Fig2}~c). For this reason is difficult to interpret our QSP sequence in terms of a physical process acting on qubits. However, we can define a physical density matrix as follows

\begin{align}
         \label{eq:QSPActionDensityMatrixPhysical}
                 \hat{\rho}=\frac{ \hat{V}_{\vec{\phi}}\hat{\Omega}_0\hat{V}^{\dagger}_{\vec{\phi}}}{\text{Tr}[\hat{V}_{\vec{\phi}}\hat{\Omega}_0\hat{V}^{\dagger}_{\vec{\phi}}]}
                     \ .
\end{align}
In this trace-preserving form, we can interpret our QSP sequence as a hybrid quantum circuit composed of unitary gates  and measurements, which is consistent with the theory of space-time dual quantum circuits~\cite{Ippoliti2021,Lu2021}.

From this discussion we can conclude that our QSP sequence with sl$(2,\mathbb{C})$ algebra does not have a finite-dimensional unitary representation. For this reason, we cannot interpret it as a unitary process acting on qubits, but rather as a hybrid process defined by Eq.~\eqref{eq:QSPActionDensityMatrixPhysical} composed by unitaries and measurements. A recent work~\cite{rossi2023} has reported a similar situation for QSP sequences using the su$(1,1)$ algebra. The associated Lie group SU$(1,1)$ also does not have a finite-dimensional unitary representation. Nevertheless, the QSP sequence can be represented as a unitary process at the expense of using continuous variables. Mathematically, this means that the group has an infinite-dimensional unitary representation~\cite{rossi2023}. Physically, this means that QSP sequences using  su$(1,1)$ algebra can be represented as operations acting on bosonic degrees of freedom. More specifically, in quantum optics, the su$(1,1)$ algebra is known to generate squeezing, which is of utmost importance in a number of fields, including quantum metrology~\cite{Guo2024} and the detection of gravitational waves~\cite{Vahlbruch2006}, as well as microwave quantum optics in superconducting devices~\cite{Moon2005}.

\section{Bosonic representation of sl$(2,\mathbb{C})$ Quantum signal processing  \label{SecV} } 
From the previous section, we have seen that there is no unitary representation of our sl$(2,\mathbb{C})$ QSP sequence. However, the algebra sl$(2,\mathbb{C})$  is the complexification of su$(1,1)$ and the group SU$(1,1)$ has a infinite-dimensional unitary representation in terms of bosonic variables~\cite{rossi2023}. Then, it is natural to think that the group SL$(2,\mathbb{C})$ also has an infinite-dimensional representation in terms of bosonic operators.

In fact, Dirac was the first to note that there exist bosonic representations of the Lorentz algebra in terms of single- and two-mode squeezing, beam splitter and phase shifter operations~\cite{Dirac1963,Bacskal2021}
\begin{align}
\label{eq:BosonicLorentz}
J_1&=\frac{1}{2}(\hat{a}^{\dagger}_1\hat{a}_2+\hat{a}^{\dagger}_2\hat{a}_1) \ ,
\nonumber\\
J_2&=\frac{1}{2\mathrm{i}}(\hat{a}^{\dagger}_1\hat{a}_2-\hat{a}^{\dagger}_2\hat{a}_1) \ ,
 \nonumber\\
  J_3&=\frac{1}{2}(\hat{a}^{\dagger}_1\hat{a}_1-\hat{a}^{\dagger}_2\hat{a}_2) \ ,
   \nonumber\\
  K_1&=-\frac{1}{4}[(\hat{a}^{\dagger}_1)^2+\hat{a}^{2}_1-(\hat{a}^{\dagger}_2)^2-\hat{a}^{2}_2] \ ,
   \nonumber\\
  K_2&=\frac{\mathrm{i}}{4}[(\hat{a}^{\dagger}_1)^2-\hat{a}^{2}_1+(\hat{a}^{\dagger}_2)^2-\hat{a}^{2}_2] \ ,
  \nonumber\\
  K_3&=\frac{1}{2}(\hat{a}^{\dagger}_1\hat{a}^{\dagger}_2+\hat{a}_1\hat{a}_2)
  \ .
\end{align}
By using these generators, we can build a unitary representation  of our sl$(2,\mathbb{C})$ QSP sequence in Eq.~\eqref{eq:LorentzQSPDef} containing a series of Gaussian operations that can be implemented in optics. 

To be more precise, in the case of sl$(2,\mathbb{C})$ QSP, we need to work in the Heisenberg picture and QSP sequences act on the bosonic operators themselves. In terms of optical operations, such a QSP sequence is composed by the combined action of a two-mode beam splitter, a phase shifter and single-mode squeezing~\cite{Weedbrook2012}. The most general action of the sequence on the bosonic operator is given by
\begin{align}
         \label{eq:QSPActionBosonic}
                  \hat{V}_{\vec{\phi}}\hat{a}_j\hat{V}^{\dagger}_{\vec{\phi}}=  A_j\hat{a}_1 + B_j\hat{a}_2  +  C_j\hat{a}^{\dagger}_1 + D_j\hat{a}^{\dagger}_2            \ .
\end{align}
The operators can be arranged in a vector form $\boldsymbol{\Psi}^{\dagger}=(\hat{a}_1,\hat{a}_2,\hat{a}^{\dagger}_1,\hat{a}^{\dagger}_2)$. Thus the QSP sequence can be written as a matrix acting on $\boldsymbol{\Psi}$ which gives us an optical input-output relation
\begin{align}
         \label{eq:QSPBosonMatrixFotm}
                  \boldsymbol{\Psi}_{\text{out}}=   \boldsymbol{V}_{\vec{\phi}} \boldsymbol{\Psi}_{\text{in}}     
                    \ .
\end{align}
Here, $\boldsymbol{V}_{\vec{\phi}} $ is the matrix representation of the QSP sequence in the Heisenberg picture.

At this stage, it is important to have a geometrical picture of our QSP sequence in the bosonic representation in mind. Note that the generators $J_j$ with $j=1,2,3$ in Eq.~\eqref{eq:BosonicLorentz} are represented in terms of beam splitters and phase shifters. This is nothing but a bosonic representation of SU$(2)$ group and these generators can be interpreted as rotations in space~\cite{Yurke1986}. They preserve the number of photons $\hat{N}=\hat{a}^{\dagger}_1\hat{a}_1+\hat{a}^{\dagger}_2\hat{a}_2$ because $[\hat{N},J_j]=0$. Geometrically, the total number of photons also defines a total angular momentum $J=N/2$ and one can think of these rotations as acting on a bloch sphere with radius $J$~\cite{Yurke1986}. This resembles the rotations of the Bloch sphere depicted in Fig.~\ref{Fig2}~c). In contrast, the generators $K_k$ with $k=1,2,3$ in Eq.~\eqref{eq:BosonicLorentz} are a bosonic representation of the Lorentz boosts in terms of single- and two-mode squeezing. They do not preserve the total number of photons~\cite{Yurke1986}. That is, if we think in terms of a Bloch sphere, these operations will deform it and they will not preserve its shape. Previously, we also noticed a similar squeezing of the Bloch sphere in Fig.~\ref{Fig2}~c) because the Lorentz boosts act as a space-time rotation creating time dilation and length contraction. From this perspective, it is natural to think that one can represent the Lorentz group using beam splitters and phase shifters to generate spatial rotations and squeezing operations generating the Lorentz boosts. Needless to say, the price we pay to have a unitary representation is that our QSP sequence cannot be naturally represented as physical operations acting on qubits, but rather on bosonic operators.

It is remarkable that quantum signal processing using sl$(2,\mathbb{C})$ algebra is related to diverse topics such space-time-dual quantum circuits generating hybrid non-unitary dynamics. Additionally, it has an infinite-dimensional unitary representation in terms of phase shifters, beams splitters, and squeezing in optics. Not to mention that is also related to the Lorentz group. Motivated by a recent work, in the next section we will discuss yet another deep connection to other field in pure and applied mathematics: The nonlinear Fourier transform~\cite{tao2012nonlinear}.

\section{Nonlinear Fourier analysis and sl$(2,\mathbb{C})$ Quantum signal processing  \label{SecVI} } 

In the preceding sections, we have discussed several aspects of quantum signal processing with  sl$(2,\mathbb{C})$ without providing any mathematical proofs of the functions that can be approximated. The reason is that, in contrast to the original QSP theorem, in our case the dynamics are non-unitary and many of the proofs require to work in the whole complex plane instead of the unit circle~\cite{Martyn2021}.

Motivated by this, in this section we adopt a distinct methodology to discuss the QSP theorem for sl$(2,\mathbb{C})$. Our strategy is to map our QSP sequence to a well-known problem in mathematics referred to as nonlinear Fourier analysis~\cite{tao2012nonlinear}. This is a mature branch of mathematics with important applications in the study of scattering phenomena~\cite{ablowitz1974}, ocean waves~\cite{osborne2018} and optical communications in nonlinear optical fibers~\cite{yousefi2014information,kamalian2016}, among others.

In this field, there are many rigorous proofs that can be adapted to QSP. In this way, we construct a solid mathematical basis for our approach.

Our strategy is based on a recent work~\cite{alexis2024}, that discusses the intriguing relation between su$(2)$ and su$(1,1)$ QSP and the nonlinear Fourier transform. Let us start by considering the QSP sequence in Eq.~\eqref{eq:ComplexQSPDef} and define
\begin{align}
         \label{eq:QSPAngles}
        \vec{\phi}&=(\phi_0,\phi_1,\ldots,\phi_{2d})
        \nonumber\\ &=
        (\psi_d,\psi_{d-1},\dots,\psi_1,\psi_0,\psi_1,\dots,\psi_{d-1},\psi_d)                  
                             \ .
\end{align}
From this we can clearly see that the QSP sequence can be written as a recurrence by defining 
\begin{align}
         \label{eq:QSPRecurrence}
                 \hat{U}^d_{\vec{\phi}}(w)=e^{\mathrm{i} \psi_d Z} e^{\mathrm{i} w X } \hat{U}^{d-1}_{\vec{\phi}}(w)  e^{\mathrm{i} w X }e^{\mathrm{i} \psi_d Z}
                    \  ,
\end{align}
where $\hat{U}^0_{\vec{\phi}}(w)=e^{\mathrm{i} \psi_0 Z}$. Amusingly, by using Lemma 1 of Ref.~\cite{alexis2024}, we can establish the intimate relation between our QSP sequence and a truncated version of the nonlinear Fourier transform for sl$(2,\mathbb{C})$~\cite{saksida2015complex,saksida2017nonlinear}
\begin{align}
         \label{eq:QSPNFT}
                 \hat{U}^d_{\vec{\phi}}(w)= e^{\mathrm{i} w X } H \hat{G}_d(z)H e^{\mathrm{i} w X }
                    \  ,
\end{align}
where $z=e^{2\mathrm{i}w}$, $H$ is a Hadamard gate such that $HZH=X$ and  $\hat{G}_d(z)$ the nonlinear Fourier series of the truncated sequence $F_n=\mathrm{i}\tan(\psi_{|n|})$ with $-d\leq n< d$. This sequence is even, purely imaginary and satisfies the relations $F_{n}=F_{-n}=-F^*_n$.

Next, let us briefly summarize the most important aspects of nonlinear Fourier analysis. Due to the symmetries of the sequence $\{F_n\}$, we can write its nonlinear Fourier transform recursively as follows
\begin{align}
         \label{eq:NLFTRecurrence}
                 \hat{G}_d(z)= \frac{1}{1-F_d^2}\begin{bmatrix}
   1& F_d e^{-\mathrm{i}dw}  \\
     F_d e^{\mathrm{i}dw} & 1  \\
    \end{bmatrix} \hat{G}_{d-1}(z)\begin{bmatrix}
   1 &  F_d e^{\mathrm{i}dw} \\
    F_d e^{-\mathrm{i}dw} & 1  \\
    \end{bmatrix}                    \  ,
\end{align}
with positive values of d and the initial condition
\begin{align}
         \label{eq:NLFTInitial}
                 \hat{G}_0(z)= \frac{1}{\sqrt{1-F_0^2}}
                 \begin{bmatrix}
   1& F_0  \\
     F_0 & 1  \\
               \end{bmatrix}                    
               \  .
\end{align}
Motivated by Ref.~\cite{alexis2024}, we use the canonical form of the nonlinear Fourier transform
\begin{align}
         \label{eq:NLFTInitial2} 
                 \hat{G}_d(z)= 
                 \begin{bmatrix}
   A(z)& B(z)  \\
    C(z)  &  D(z)  \\
               \end{bmatrix}                    
               \  .
\end{align}
such that $A(z)D(z)+B(z)C(z)=1$. By replacing this in Eq.~\eqref{eq:QSPNFT}, we obtain a relation between the functions defining the nonlinear Fourier transformation and the QSP sequence.

\begin{figure}
\centering
    \includegraphics[width=0.40 \textwidth]{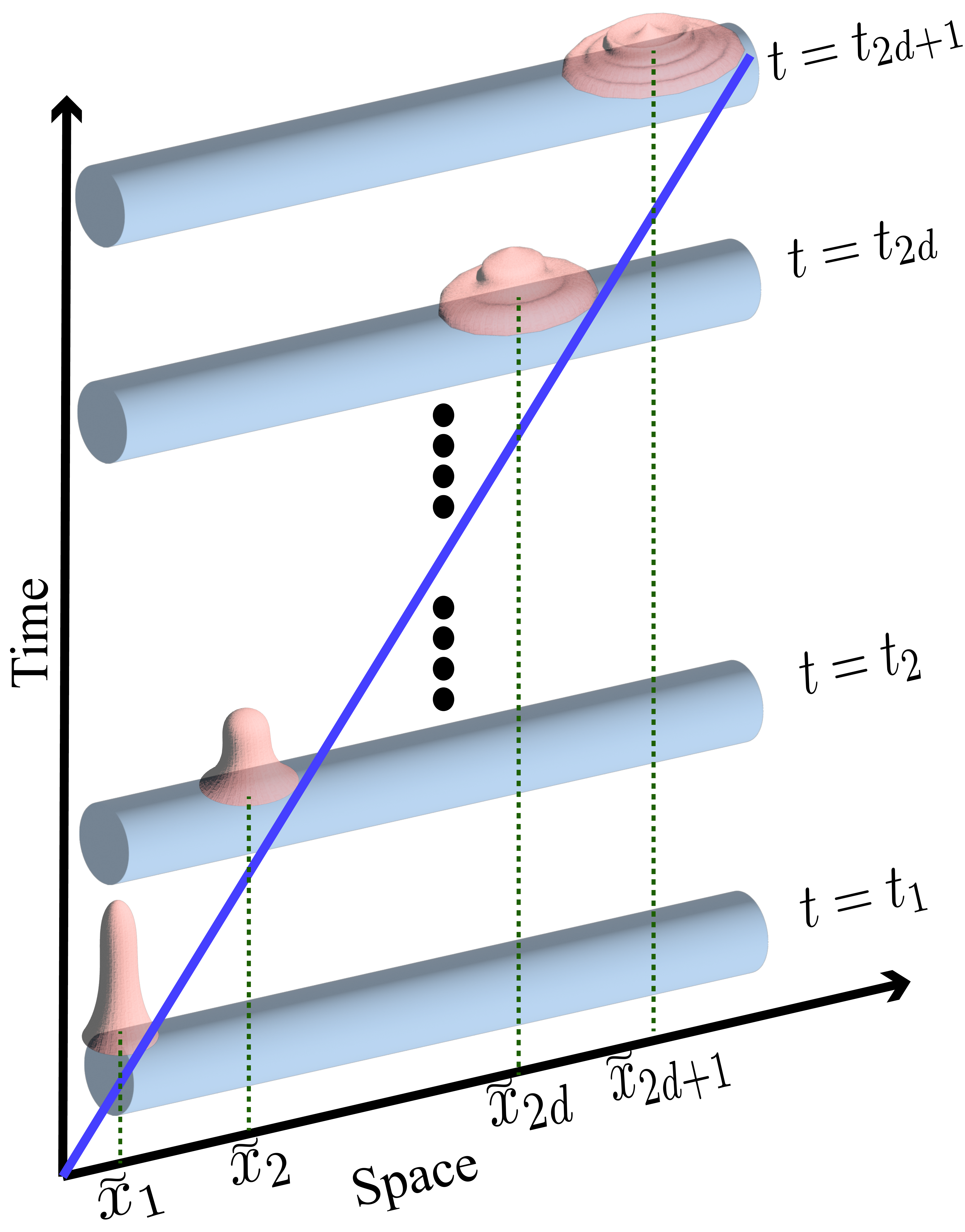}
    \caption{Illustration of the relation between propagation of nonlinear waves, nonlinear Fourier analysis and quantum signal processing.  An initial wave form $f(x,0)=F(x)$
    propagates through a nonlinear media and it changes its shape at different positions $\tilde{x}_n$ (with $\tilde{x}_1=x_{-d}$ and $\tilde{x}_{2d+1}=x_{d}$) and times $t_n$ (with $t_1=0$). The nonlinear propagation equation $\partial_t f(x,t)=K[f(x,t)]$ where $K(z)$ is a nonlinear function, can be encoded into two linear differential equations for an auxiliary wave $\psi(x,t)$. One of them for time, and another for space propagation (see appendix~\ref{AppendixA}). The nonlinear Fourier transform is determined by solving the spatial propagation equation $\partial_x \psi(x,0)=\boldsymbol{A}(0)\psi(x,0)$, where $\boldsymbol{A}(0)$ belongs to the sl$(2,\mathbb{C})$ algebra. The nonlinear wave equation determines the Lie-group used in nonlinear Fourier analysis and the associated QSP sequence.}
    \label{Fig4}
\end{figure}

Next, we will discuss the physical meaning of the relation between QSP with sl$(2,\mathbb{C})$ and the nonlinear Fourier transform. Figure~\ref{Fig4} depicts the propagation of a wave $f(x,t)$ in a nonlinear media.
Ideally, in the continuous limit, the propagation of the wave may take place between in the whole real line $-\infty<x<\infty$ with an initial condition $f(x,0)=F(x)$ at time $t_1=0$. As we depict in Fig.~\ref{Fig4}, the wave will propagate through the nonlinear media and change its shape during the evolution. Due to the nature of the media, the evolution of the wave is governed by a partial differential equation $\partial_t f(x,t)=K[f(x,t)]$, where $K[f(x,t)]$ is a nonlinear function of the wave $f(x,t)$. Using the tools discussed in appendix~\ref{AppendixA}, one can encode the nonlinear equation for $f(x,t)$ into two linear equations
\begin{align} 
\label{eq:CompatibilityConditionsMain2} 
 \partial_t \psi(x,t)=\boldsymbol{P}(t)\psi(x,t)   , \ \partial_x \psi(x,t)=\boldsymbol{A}(t)\psi(x,t)
 \end{align}
with suitable differential operators $\boldsymbol{P}(t)$ and $\boldsymbol{A}(t)$, and an auxiliary wave $\psi(x,t)$. Now, if we forget about time and set $t=0$, we can propagate $ \psi(x,0)$ in space by solving the equation $\partial_x \psi(x,0)=\boldsymbol{A}(0)\psi(x,0)$. Of course, the nonlinear wave $f(x,t)$  propagates in space and time, but the auxiliary wave $ \psi(x,t)$ can propagate independently in each direction.

Next, let us consider a family differential operators $\boldsymbol{L}(t)$ that are isospectral, i.e., have the same eigenvalues $\lambda(t)=\lambda$ where $\boldsymbol{L}(t)\psi(x,t)=\lambda\psi(x,t)$  for different values of $t$.  To propagate the auxiliary wave $\psi(x,t)$ in space, we work in a rotating frame where the operator $\boldsymbol{A}(t)$ takes the form (see appendix~\ref{AppendixA})
\begin{align}
         \label{eq:MainRotatingOperatorPNLS}
                  \boldsymbol{A}_{\text{g}}&=\begin{bmatrix}
   0 & r(x,t)e^{-\mathrm{i}\lambda x}  \\
    s(x,t)e^{\mathrm{i}\lambda x}  &  0  \\
    \end{bmatrix}
    \ ,
\end{align}
$r(x,t)$ and $s(x,t)$ are related to the wave $f(x,t)$.
If we define a  propagator $\mathcal{G}(x,t)$ for the compatibility conditions in Eq.~\eqref{eq:CompatibilityConditionsMain2} such that $\psi(x,t)=\mathcal{G}(x,t)\psi(-\infty,0)$, one can show that $\partial_x \mathcal{G}(x,t)=\boldsymbol{A}_{\text{g}}(t)\mathcal{G}(x,t) $.

Now we have all the basic elements we require to relate this discussion to the nonlinear Fourier transformation $ \hat{G}_d(z)$ in \eqref{eq:NLFTRecurrence}. To establish this relation, we need to discretize space defining $x_n=n\Delta$ with integer $n=-d,-d+1,\ldots, 0,d-1,d$ and $\Delta$ being small and replace the wave $f(x,0)$ by a discrete sequence $F_n=f(x_n,0)$. By using this, we can discretize the compatibility condition in the rotating frame $\partial_x \mathcal{G}(x,0)=\boldsymbol{A}_{\text{g}}(0)\mathcal{G}(x,0) $, as follows
\begin{align}
         \label{eq:Discretization1}
                  \frac{\mathcal{G}_{n+1}-\mathcal{G}_n}{\Delta}=\begin{bmatrix}
   0 & r_n e^{-\mathrm{i}\lambda \Delta n}  \\
   s_n e^{\mathrm{i}\lambda  \Delta n}  &  0  \\
    \end{bmatrix}\mathcal{G}_n
                  \ .
\end{align}
where $\mathcal{G}_n= \mathcal{G}(x_{n},0)$, $r_n=r(x_n,0)$  and $s_n=s(x_n,0)$.   From this we can obtain the recurrence relation
\begin{align}
         \label{eq:Discretization2}
                 \mathcal{G}_{n+1}=\begin{bmatrix}
   1& \Delta r_n e^{-\mathrm{i}\lambda  \Delta n}  \\
  \Delta s_n e^{\mathrm{i}\lambda  \Delta n}  &  1  \\
    \end{bmatrix}\mathcal{G}_{n}
                  \ .
\end{align}
A crucial aspect in our derivation is to define $\Delta r_n=F_ne^{\eta n/2}$, $\Delta s_n=F_ne^{-\eta n/2}$ and $\lambda\Delta=\delta/2$ to obtain the recurrence 
\begin{align}
         \label{eq:Discretization3}
                 \hat{G}_{n+1}(z)=\frac{1}{\sqrt{1-F_n^2}}\begin{bmatrix}
   1& F_n e^{-\mathrm{i}n w }  \\
 F_n e^{\mathrm{i}n w }  &  1  \\
    \end{bmatrix}  \hat{G}_{n}(z)
\end{align}
with $w=(\delta+\mathrm{i} \eta)/2$. Note that both $\hat{G}_{n}(z)$ and $\hat{G}_{n+1}(z)$ are elements of the Lie group SL$(2,\mathbb{C})$ and $\det\hat{G}_{n}(z)=\det \hat{G}_{n+1}(z)=1$. For the discretization to be consistent with the group structure, we had to add an extra factor to the recurrence Eq.~\eqref{eq:Discretization3}.

Next, let us propagate the system starting from $n=-d$ to $n=d$ corresponding to spatial propagation between the discrete coordinates $x_{-d}$ and $x_{d}$ under the assumption that $ \hat{G}_{-d}(z)=\hat{1}$. As we did before, if we further assume that sequence $\{F_n\}$ is even, purely imaginary and satisfies the relations $F_{n}=F_{-n}=-F^*_n$, one obtains Eq.~\eqref{eq:NLFTRecurrence}.

We have learned several things from this derivation. The first and most important one is that the sequence $F_n=f(x_n,0)$, is related to a wave that in the continuous limit that satisfies a non-linear differential equation. This wave defines two functions $r_n$ and $s_n$. In the context of QSP, the sequence $F_n$ is directly associated to the QSP sequence in Eq.~\eqref{eq:QSPAngles} via the relation $F_n=\mathrm{i}\tan(\psi_{|n|})$ with $-d\leq n< d$.  The signal $w$ is the parameter that defines the variable $z=e^{2\mathrm{i}w}$ in the nonlinear Fourier transform and is directly associated to the spectrum of the differential operator $\boldsymbol{L}(t)$. We refer the reader to  appendix~\ref{AppendixA} for detailed information.

\section{Conclusions  \label{Conclusions} } 
In summary, motivated by the relation between space-time dual quantum circuits and complexification of QSP for the su$(2)$ algebra, we examined in detail the possible relations and ramifications of this idea to other related concepts and fields. The core mathematical structure of these relations is the Lie algebra sl$(2,\mathbb{C})$, which is a complexification of su$(2)$. In its non-unitary finite-dimensional representation, our QSP sequence sl$(2,\mathbb{C})$ can be interpreted in terms of the Lorentz transformations acting on density matrices. As the Lorentz transformations do not preserve the trace in this representation, the QSP sequence can be interpreted as a hybrid quantum process composed by unitaries and measurements as in Refs.~\cite{Ippoliti2021,Lu2021}. Besides that, we showed that there is a unitary infinite-dimensional representation of our QSP sequence given in terms of bosonic two-mode squeezing. Further, we discussed the relation to nonlinear Fourier analysis for the group SL$(2,\mathbb{C})$~\cite{saksida2015complex,saksida2017nonlinear}, which extends the recent work relating su$(2)$ and su$(1,1)$ QSP sequences to the nonlinear Fourier transform~\cite{alexis2024}.

It would be interesting to explore applications of our approach to extend concepts such as Quantum Singular Value transformation (QSVT), qubitization and block encodings within our framework of complexified QSP. One possible direction of research is to further explore the non-unitay representation of our QSP sequence and interpretet it as a hybrid QSP involving unitary evolution and measurements. This can lead to interesting questions such as purification and measurement-induced phase transitions Refs.~\cite{Ippoliti2021,Lu2021}, which are of interest for the community. Another possible extension of our work is to explore the relativistic character of our approach and its effect on quantum states. It is definitely interesting to further investigate the mathematical foundations of the relation between complexified QSP and nonlinear Fourier analysis for the group SL$(2,\mathbb{C})$~\cite{saksida2015complex,saksida2017nonlinear}. On top of that, this intriguing relation can potentially be applied to simulate scattering and nonlinear wave propagation phenomena in quantum computers. In future works, it would be interesting to exploit methods of QSP to make statements of interest for pure mathematics. In appendix~\ref{AppendixB}, we show that one can map our QSP sequence in terms of M\"obius transformations. This could open a new perspective on mathematical questions such as conformal maps~\cite{MunizOliva2002}.
We emphasize that this work opens several directions of research and allows more complex applications that leverage QSP as a powerful tool.

\textit{Acknowledgments ---} The authors would like to thank I. L. Chuang, T. Holdsworth, J. M. Martyn, and W. E. Salazar for valuable discussions and NTT Research
Inc.  for their support in this collaboration. 

\appendix

\section{Complex Nonlinear Fourier transform, integrability and the  sl$(2,\mathbb{C})$ algebra \label{AppendixA} } 
In the main text we discussed the relation between quantum signal processing for sl$(2,\mathbb{C})$ algebra and the complex nonlinear Fourier transform. This is a very deep and highly nontrivial connection. For this reason, we discuss in detail how these two concepts are related. It is evident that the most prudent course of action would be to start with a concise overview of the nonlinear Fourier transformation and its relationship to the integrability of specific types of nonlinear differential equations. This appendix is mostly based on the discussion presented in Ref.~~\cite{yousefi2014information}, so we refer the reader to the original reference for further details.

Let us start by briefly discussing integrability and Lax pairs. The first step is to define a wave $f(x,t)$ defining a field that is a function of space $x$ and time $t$ coordinates. The wave $f(x,t)$ can be used to construct a linear differential operator whose evolution preserves desired spectral properties of a linear operator. To formalize the discussion, let us assume that there is a matrix $\boldsymbol{L}(t)$ defining a linear differential operator. Let us assume that this matrix is diagonalizable by a similarity transformation $\boldsymbol{O}(t)$ such that $\boldsymbol{L}(t)=\boldsymbol{O}(t)\boldsymbol{L}_0\boldsymbol{O}^{-1}(t)$, where $\boldsymbol{L}_0=\boldsymbol{L}(0)$ sets the initial condition. Clearly this  family of operators depending on $t$ and we assume that its dependence on $t$ is smooth. One may ask what are the conditions for which the family matrices $\boldsymbol{L}(t)$ are isospectral, i.e., have the same eigenvalues $\lambda(t)=\lambda$ where $\boldsymbol{L}(t)\psi(x,t)=\lambda\psi(x,t)$  for different values of $t$. To answer this question, it is useful to calculate the derivative $\boldsymbol{L}(t)$ with respect to $t$, which defines the Lax equation, as follows

\begin{align}
\label{eq:Isospectral}
\frac{d \boldsymbol{L}(t)}{dt}=[\boldsymbol{P}(t),\boldsymbol{L}(t)]
\ ,
\end{align}
where $\boldsymbol{O}(t)$ is a solution of the differential equation $\frac{d \boldsymbol{O}(t)}{dt}=P(t)\boldsymbol{O}(t)$ and $[\boldsymbol{P}(t),\boldsymbol{L}(t)]=\boldsymbol{P}(t)\boldsymbol{L}(t)-\boldsymbol{L}(t)\boldsymbol{P}(t)$ is the usual commutator. In fact, if $\boldsymbol{L}(t)$ is diagonalizable, it has to satisfy Eq.~\eqref{eq:Isospectral}.

Next let us assume that the matrices  $\boldsymbol{L}(t)$ and $\boldsymbol{P}(t)$ are not independent but they are functions of the wave  $f(x,t)$. Note that, for the purposes of this work, and to avoid confusion with the quantities used in Quantum Signal Processing (QSP), we refer to $f(x,t)$ as a wave instead of the signal, as it is usually done in the literature on non-linear Fourier transform.

The matrices $\boldsymbol{L}(t)$ and $\boldsymbol{P}(t)$ are referred to as a Lax pair in the literature.

By exploiting the isospectral property of Eq.~\eqref{eq:Isospectral}, one can define non-linear evolution equations for the wave of the form
\begin{align}
\label{eq:NonlinearEq}
\partial_t f(x,t)=K[f(x,t)]
\ ,
\end{align}
where $K[f(x,t)]$ is a nonlinear function of the wave $f(x,t)$. At this stage it is useful to consider an example. Let us define the linear operators $\boldsymbol{L}(t)=\partial^2_x+f(x,t)$ and $\boldsymbol{P}(t)=\frac{3}{2}\partial^3_x+\frac{1}{2}(\partial_x f(x,t)+f(x,t)\partial_x)$. The Lax equation Eq.~\eqref{eq:Isospectral} giving rise to an isospectral family of operators $\boldsymbol{L}(t)$ implies that the wave $f(x,t)$ is a solution of the a nonlinear differential equation
\begin{align}
\label{eq:KDV}
\partial_t f(x,t)=\partial^3_x f(x,t)+f(x,t)\partial_x f(x,t)
\ ,
\end{align}
which is known in the literature as the Korteweg–De Vries (KdV) equation that governs the dynamics of nonlinear waves in shallow water surfaces.

Next, let us consider the eigenvalue equation $\boldsymbol{L}(t)\psi(x,t)=\lambda \psi(x,t)$. If we take the derivative with respect to $t$ and use the Lax equation, we can obtain the linear equations 
\begin{align} 
\label{eq:CompatibilityConditions}
 \partial_t \psi(x,t)=\boldsymbol{P}(t)\psi(x,t)   , \ \partial_x \psi(x,t)=\boldsymbol{A}(t)\psi(x,t)
 \end{align}
for some linear differential operator $\boldsymbol{A}(t)$ that is related to $\boldsymbol{L}(t)$ via the identity

\begin{align}
\label{eq:Conection}
\boldsymbol{A}(t)=\mathrm{i}\sigma_z(\boldsymbol{L}(t)-\hat{\boldsymbol{1}})+\partial_x
\ ,
\end{align}
where $\sigma_z$ is the usual Pauli matrix and $\hat{\boldsymbol{1}}$ is the identity in the operator space.
Next, by using the identity $\partial_x\partial_t \psi(x,t)=\partial_t\partial_x \psi(x,t)$, one can show that the Lax equation can be rewritten in terms of the operator $\boldsymbol{A}$, as follows
\begin{align}
\label{eq:CurvatureCondition}
\partial_t\boldsymbol{A}(t)-\partial_x\boldsymbol{P}(t)+[\boldsymbol{A}(t),\boldsymbol{P}(t)]=0
\ ,
\end{align}
which known in the literature as the zero-curvature condition. One remarkable aspect is that the operator $\boldsymbol{A}(t)$ defines a non-linear differential equation for the wave $f(x,t)$, and this nonlinear behavior arises from the compatibility between two linear equations in Eq.~\eqref{eq:CompatibilityConditions}.

Previous works on the nonlinear Schr\"odinger equation by Zakharov and Shabat, discussed the canonical form of the operator $\boldsymbol{A}(t)$, as follows
 
\begin{align}
         \label{eq:OperatorPNLS}
                  \boldsymbol{A}(t)=\begin{bmatrix}
   \mathrm{i}\lambda & r(t,x)  \\
    s(t,x) &  -\mathrm{i}\lambda   \\
    \end{bmatrix}= \mathrm{i}\lambda\sigma_z+r(t,x) \sigma^+ + s(t,x) \sigma^-
  \ ,
\end{align}
where $2\sigma^{\pm}=\sigma_x\pm\mathrm{i}\sigma_y$ are the ladder operators of the su$(2)$ algebra. The functions $r(x,t)$ and $s(x,t)$ are intimately related to the wave $f(x,t)$, as we will see later. It is important to see the explicit form of the operator $\boldsymbol{L}(t)$ appearing in the Lax equation~\eqref{eq:Isospectral}, which can be directly obtained from Eq.~\eqref{eq:Conection} and reads

\begin{align}
         \label{eq:OperatorLNLS}
                  \boldsymbol{L}(t)=\mathrm{i}\begin{bmatrix}
   \partial_x & -r(x,t)  \\
    s(x,t) &   -\partial_x   \\
    \end{bmatrix}=\mathrm{i} \sigma_z \partial_x-\mathrm{i}r(x,t) \sigma^++\mathrm{i}s(x,t) \sigma^-
  \ .
\end{align}
For the operator $ \boldsymbol{A}(t)$ in the canonical form of Eq.~\eqref{eq:OperatorPNLS}  the differential equation  $\partial_x \psi(x,t)=\boldsymbol{A}(t)\psi(x,t)$  is known in the literature as the AKNS system (AKNS stands for the authors M. J. Ablowitz, D. J. Kaup, A. C. Newell, and H. Segur of the original paper~\cite{ablowitz1974}). 

At this point of the discussion, it is important to give an example. Let us consider the important case $r(x,t)=f(x,t)$ and $s(x,t)=-f^*(x,t)$ that is generally known as the Zakharov-Shabat system. For this example, the algebra associated to the operator
 
\begin{align}
         \label{eq:ExampleOperatorPNLS}
                  \boldsymbol{A}(t)=\begin{bmatrix}
   \mathrm{i}\lambda & f(x,t)  \\
   -f^*(x,t) &  -\mathrm{i}\lambda   \\
    \end{bmatrix}
  \ ,
\end{align}
 is su$(2)$. The matrix $\boldsymbol{P}(t)$ is given explicitly by
\begin{align}
         \label{eq:ExplicitLNLS}
                  \boldsymbol{P}(t)=\begin{bmatrix}
   \mathrm{i}(2\lambda^2-|f(x,t)|^2) & 2\lambda f(x,t)- \mathrm{i} \partial_x f(x,t) \\
    -2\lambda f^*(x,t)- \mathrm{i} \partial_x f^*(x,t) &   -\mathrm{i}(2\lambda^2-|f(x,t)|^2)   \\
    \end{bmatrix}
  \ .
\end{align}

By using this, to satisfy the zero-curvature condition in Eq.~\eqref{eq:CurvatureCondition}, the wave $f(x,t)$ has to satisfy the non-linear Schr\"odinger equation
\begin{align}
         \label{eq:NLS}
                   \mathrm{i}\partial_t f(x,t)=\partial^2_x f(x,t)+2f(x,t)|f(x,t)|^2
                    \ .
\end{align}

So far, we have summarized the most important aspects of integrability and the Lax pairs. However, it is not so clear how this discussion is related to quantum signal processing and the nonlinear Fourier transformation. To establish this deep relation between these concepts, it is convenient to start from the equation $\partial_x \psi(x,t)=\boldsymbol{A}(t)\psi(x,t)$ and work in a rotating frame such that the diagonal term of the operator $\boldsymbol{A}(t)$ in Eq.~\eqref{eq:OperatorPNLS} in that frame is effectively eliminated. This can be achieved by defining a gauge transformation $\psi(x,t)=g(x)\psi_{\text{g}}(x,t)$. The equation of motion for the rotating spinor $\psi_{\text{g}}(x,t)$ is given by $\partial_x \psi_{\text{g}}(x,t)=\boldsymbol{A}_{\text{g}}(t)\psi_g(x,t)$, thus defining the operator in the rotating frame $\boldsymbol{A}_{\text{g}}(t)=-g^{-1}(x)\partial_xg(x)+g^{-1}(x)\boldsymbol{A}(t)g(x)$. By choosing $g(x)=e^{\mathrm{i}\lambda\sigma_z x}$, we obtain

\begin{align}
         \label{eq:RotatingOperatorPNLS}
                  \boldsymbol{A}_{\text{g}}&=\begin{bmatrix}
   0 & r(x,t)e^{-\mathrm{i}\lambda x}  \\
    s(x,t)e^{\mathrm{i}\lambda x}  &  0  \\
    \end{bmatrix}
    \nonumber \\
    &= r(x,t)e^{-\mathrm{i}\lambda x}  \sigma^+ + s(x,t) \sigma^-e^{\mathrm{i}\lambda x} 
  \ .
\end{align}
As $\text{Tr}[A_{\text{g}}(t)]=0$, we can ensure that $ \boldsymbol{A}_{\text{g}}(t)$ is an element of the algebra  sl$(2,\mathbb{C})$.
It us useful to define a  propagator $\mathcal{G}(x,t)$ for the compatibility conditions such that $\psi(x,t)=\mathcal{G}(x,t)\psi(-\infty,0)$. For convenience, we define the spatial coordinate in the range $-\infty<x<\infty$. In our definition of the propagator, we consider the initial spatial coordinate $x_1=-\infty$ and initial time $t_1=0$.  One can show that the propagator also solves the compatibility conditions $\partial_x \mathcal{G}(x,t)=\boldsymbol{A}_{\text{g}}(t)\mathcal{G}(x,t) $ of Eq.~\eqref{eq:CompatibilityConditions}, which has a formal solution
\begin{align}
         \label{eq:Propagator1}
                  \mathcal{G}(x,t)= \mathcal{G}(-\infty,t)+\int^x_{-\infty} \boldsymbol{A}_{\text{g}}(t)\mathcal{G}(x',t) dx'
                  \ ,
\end{align}
and is an element of the group SL$(2,\mathbb{C})$ discussed in this paper. Of particular interest is the case $t=0$, where this equation reads

\begin{align}
         \label{eq:Propagator2}
                  \mathcal{G}(x,0)=\hat{1}+\int^x_{-\infty} \boldsymbol{A}_{\text{g}}(0)\mathcal{G}(x',0) dx'
                  \ .
\end{align}
Here we have used that  $\mathcal{G}(-\infty,0)=\hat{1}$. The nonlinear Fourier transform is defined as $\mathcal{G}(\infty,0)$.

\section{Non-unitary Quantum signal processing and M\"obius transformations \label{AppendixB} } 
In the previous sections, we discussed several representations of sl$(2,\mathbb{C})$ QSP. In this section, we consider a rather abstract representation in terms of the M\"obius transformations. Let us consider the action of the QSP sequence in Eq.~\eqref{eq:ComplexQSPDef} on a qubit
\begin{align}
\label{eq:ComplexQSPQubit}
\begin{bmatrix}
    \psi'_1  \\
     \psi'_2  \\
    \end{bmatrix} =   
    \begin{bmatrix}
    P(\cos w) &  \mathrm{i} Q(\cos w) \sin w  \\
     \mathrm{i}R(\cos w) \sin w &S(\cos w)  \\
    \end{bmatrix}\cdot \begin{bmatrix}
    \psi_1  \\
     \psi_2  \\
    \end{bmatrix}
\ .
\end{align}
As this matrix is in general non-unitary, it does not preserve the norm of the quantum state and even if the initial state is normalized, $|\psi_1|^2+|\psi_2|^2 = 1$, the final state is not. For this reason, when the matrix is non-unitary what matters is the scale factor $z'= \psi'_1/ \psi'_2$ and its evolution. More explicitly, we can represent QSP as an evolution equation for scale factors of the probability amplitudes, as follows
\begin{align}
\label{eq:QSPMoebius}
z'&=\frac{\psi'_1}{\psi'_2}=\frac{P(\cos w) \psi_1+ \mathrm{i} Q(\cos w) \sin w \psi_2}{ [\mathrm{i}R(\cos w) \sin w]\psi_1+S(\cos w)\psi_2}
\nonumber\\
&=\frac{P(\cos w) z+ \mathrm{i} Q(\cos w) \sin w }{ [\mathrm{i}R(\cos w) \sin w] z+S(\cos w)}
\ ,
\end{align}
where $z= \psi_1/ \psi_2$.
This has exactly the canonical form of a M\"obius transformation
\begin{align}
\label{eq:CanonicalMoebius}
z'=\frac{a z+ b}{ c z+d}
\ ,
\end{align}
where $ab-cd=1$. Any  M\"obius transformation can be obtained from the composition of two automorphisms $1/z$ and $ez+f$ with $e\neq 0$~\cite{MunizOliva2002}.

\end{document}